\documentclass[12pt,preprint]{aastex}

\shorttitle{Near moving groups}
\shortauthors{L\'opez-Santiago et al.}

\begin{document}


\title{The nearest young moving groups}


\author{J. L\'opez-Santiago\altaffilmark{1,2}, D. Montes\altaffilmark{2},
I. Crespo-Chac\'on\altaffilmark{2} and M.J. Fern\'andez-Figueroa\altaffilmark{2}}
\affil{\altaffilmark{1}INAF - Osservatorio Astronomico di Palermo,
Piazza Parlamento 1, I-90134 Palermo, Italy}
\affil{\altaffilmark{2}Departamento de Astrof\'{\i}sica y Ciencias de la
Atm\'osfera, Facultad de Ciencias F\'{\i}sicas,
Universidad Complutense de Madrid, E-28040 Madrid, Spain}




\begin{abstract}
The latest results in the research of forming planetary systems
have led several authors to compile a sample of candidates for searching
for planets in the vicinity of the sun. Young stellar associations
are indeed excellent laboratories for this study, but some of them
are not close enough to allow the detection of planets through
adaptive optics techniques.
However, the existence of very close young moving groups can solve this
problem.
Here we have compiled the members of the nearest young moving groups,
as well as a list of new candidates from our catalogue of {\it late-type stars
possible members of young stellar kinematic groups}, studying their
membership through spectroscopic and photometric criteria.
%
\end{abstract}





\section{Introduction}
\label{sec:intr}

In recent years, a series of young stellar kinematic groups
(clusters, associations, and moving groups) of late-type stars
with similar space motion and ages ranging from 8 to 50 Myr
\citep[see][and references therein]{zuc04a}
has been discovered in our neighbourhood:
TW~Hya, $\beta$~Pic, AB~Dor, $\eta$~Cha, $\epsilon$~Cha, Tucana and
Horologium associations. In addition, several more distant young
associations such as MBM~12 \citep{hea00}, Corona Australis \citep{qua01},
and possibly the group of stars with a motion similar to that of HD~141569
\citep{wei00} have also been identified.
In the Galactic velocity space, they situate inside the boundaries of the
\object{Local Association} (see Fig.~\ref{fig:uv}), a mixture of young stellar
complexes ---~OB and T-associations~--- and clusters with different
ages \citep{egg75, egg83a, egg83b, m01}.
%
These associations of very young stars are excellent laboratories for
investigations of forming planetary systems \citep{zuc04}.
Nevertheless, they are generally situated at distances above 50 pc,
which makes them less accessible to adaptive optics systems even on large
telescopes.

It is well-known that tightly bound, long-lived open clusters can account
for only a few per cent of the total galactic star formation rate
\citep{wie71}. Therefore, either most clusters and associations
disperse very quickly after star formation has started or most are
born in isolation \citep{wic03}.
The existence of very young moving groups (MGs) with a few dozens of stars
showing the same spectroscopic properties ---~i.e. age, metallicity, level
of magnetic activity~--- is in agreement with the first explanation.
Small associations of stars may be dispersed by galactic differential rotation
since they are not gravitationally bounded enough, taking into
account that their nucleus consist of only a few stars as in the case
of the Ursa Major MG (see King et al. 2003, for a recent review) or
the recently discovered AB~Doradus MG \citep{zuc04}.
The location of these young MGs inside the Local Association and its
proximity in the $UV$-plane can be explained as the result of
the juxtaposition of several star forming bursts in adjacent cells of the
velocity field (see Montes et al. 2001, and references therein) or
dynamical perturbations caused by spiral waves \citep{sim04,fam05,qui05}.
Thus, one expects to find groups of coeval stars with similar space motion in
our neighbourhood.
%

In 2001, the $\beta$~Pic~MG \citep{zuc01} --- a group of stars with an age of
$\sim$~12~Myr \citep{zuc01, ort04} at a mean distance of $\sim$~35~pc
co-moving with the well-known young star $\beta$~Pic --- was confirmed
to be the closest kinematic group up to date.
More recently, \citet{zuc04} have
identified a new group of stars co-moving with the also well-known young star
AB~Dor, at a mean distance of $\sim$~30~pc, and with an age of $\sim$~50~Myr.
Nevertheless, the existence of a nearer association of a few stars was
proposed by \citet{gai98} and studied in detail by \citet{fur04}, though
its existence is quite controversial.
Here we discuss about the fact of the nearest MGs using both spectroscopic
and photometric criteria of membership
for a sample of stars that includes the proposed members
from the literature and our list of young cool stars possible members of
young stellar kinematic groups
\citep{m01, lop05}.

\section{The Hercules-Lyra Association}
\label{sec:herc}

Based on the kinematics of young solar analogues in the solar neighbourhood,
\citet{gai98} confirmed the existence of a group of four stars
(marked with $\dag$ in Table~\ref{tab:MGs}) co-moving in the space towards the
constellation of Hercules. Recently, \citet{fur04} has extended the sample of
late-type stars of this MG up to 15 nearby (d~$<$~25~pc) candidates,
proposing the name Hercules-Lyra since several members show a radiant
``{\it evenly matched}'' with this constellation.
Comparing the level of chromospheric activity of the stars of his sample
with that of the members of the Ursa Major Association and looking for
the existence of lithium in their spectrum, he notices that
several candidates of Hercules-Lyra appear to be coeval of the Ursa Major
stars, for which he gives an age of $\sim$~200~Myr. On the contrary,
other candidates seem to be older (e.g. HD~111395) or younger (HD~17925,
HD~82443, and HD~113449), questioning the existence of Hercules-Lyra as an
entity independent of the Local Association. However, he considers unlikely
that the majority of his sample can originate from the Pleiades alone, or
other clusters of the Local Association since ``{\it they are poorer and more
distant}'' as pointed out by \citet{jef95}. Thus, he confirms
``{\it the bulk}'' of the sample --- formed by the stars HD~166, HD~96064,
HD~97334, HD~116956, HD~139777, HD~139813 and HD~141272 (see his Table~1) ---
to be an entity on its own.

Here we discuss the possible existence of the Hercules-Lyra MG
as an independent association using kinematic (space motion), spectroscopic
(lithium abundance) and photometric (isochrone fitting) criteria. A total of
12 possible members (stars marked with {\it a} in Table~\ref{tab:MGs})
have been added to the initial sample of \citet{fur04} from our
catalogue of {\it Late-type Stars Possible Members of Young Stellar
Kinematic Groups} \citep{m01}.
The candidates have been chosen by their kinematics assuming a total
dispersion of \mbox{$\pm$ 6 km~s$^{-1}$} in $U$ and $V$, respectively;
that is, an average position of \mbox{($U$, $V$) = (-15.4, -23.4)}
km~s$^{-1}$
has been determined using the stars given by \citet{fur04}, and every star
in our catalogue in a radius of \mbox{$\pm$ 6 km~s$^{-1}$} has been
selected. The value of the dispersion has been chosen equal to that of the
\mbox{$\sim 200$ Myrs} old Castor~MG \citep{m01}, coeval of the
Hercules-Lyra Association. No restriction in the $W$ component has been
imposed in this first selection.

In Table~\ref{tab:MGs} we summarize the results obtained by us.
From the whole sample of 27 candidates, eight stars have been discarded
as members by their space motion: HD~25457, located inside the B4
subgroup (see Fig.~\ref{fig:uv}); HD~96064, HD~112733, HIP~67092, the
binary system made up of the F-type star HD~139777 and HD~139813, and
HD~207129 all them with a value in $W$ higher than that of the rest of the
candidates (see Fig.~\ref{fig:uvw_MGs}); and HD~113449, classified as
member of the AB~Dor~MG by \citet{zuc04} (see Table~\ref{tab:MGs} and
$\S$~\ref{sec:abdor} for a more detailed discussion) and
questioned by \citet{fur04} because of its relatively high lithium abundance.
We have also studied the lithium abundance --- measured as the equivalent width
of the lithium line $\lambda$6707.8~\AA, $EW$({Li~{\sc i}}) --- in each one
of the candidates. The values of $EW$(Li~{\sc i}) have been taken from
\citet{lop05} and compared with those of the members of well-known stellar
clusters (see Fig.~\ref{fig:li_MGs}). The results appear to be consistent
with an age of \mbox{150 -- 300~Myr} for seven candidates. However, several
stars (HD~1466, HD~17295, \mbox{1E~0318.5-19.4}, and HD~82443) show an
$EW$(Li~{\sc i}) comparable to that of the members of the Pleiades while
other five (HD~37394, HD~97334B, HD~111395, HD~116956 and HD~141272) are fully
depleted or have a value lower than the expected for a member of the Hercules-Lyra
Association.
For isochrone fitting, we have adopted pre-main sequence models
from \citet{sie00}. For $T_{\rm eff} < 4000$~K,
the models systematically  underestimate the age when comparing
with clusters of known age such as the Pleiades and IC~2391
in a $M_{\rm v}$ vs. $V-I$ diagram \citep{lop05} due to the transformation
from flux to colour. Bearing this in mind, the corrected transformation
adopted by \citet{lop03} and \citet{lop05b}
for stars cooler than 4000~K has been used in this work.
The values of $V-I$ have been taken from the Hypparcos Catalogue \citep{esa97}.
The result of comparing the position of the stars with the isochrones
in the colour-magnitude diagram (CMD) (Fig.~\ref{fig:VI}) is again in agreement
with and age of \mbox{$\sim$ 150 - 300 Myrs}. Nevertheless, no conclusions
can be inferred from the CMD alone since isochrones of more than 80~Myr
converge for $V-I \le 1.8$ mag., and ages larger than 300 Myr could be
adopted.

From the combination of the three criteria, the total sample of candidates is
reduced to 10 stars with $EW$(Li~{\sc i}) and position in the CMD compatibles
with an age of \mbox{$\sim$ 200 Myrs}, which could form the bulk of the
Hercules-Lyra Association, and other 15 definitively non members or with a
doubtful classification (Table~\ref{tab:MGs}). The members show a deviation
\mbox{($\sigma_{\rm U}$, $\sigma_{\rm V}$) = (2.46, 1.61) km s$^{-1}$} from
the centre (\mbox{($U$, $V$) = (13.19, 20.64) km s$^{-1}$}) lower than
that of other coeval MGs such as Castor and Ursa Major
\citep[see][and references therein]{m01}.
%
A similar dispersion (\mbox{$\sigma_{\rm W} \approx 3.4$ km s$^{-1}$}) is
found in W, confirming the results in $U$ and $V$. In the same way, the shape of
the MG in the velocity field is in agreement with the theory of the MGs
\citep{egg65, age79, sku99, asi99b, lop05}. According to this theory, not-gravitational
bounded stars formed in the same forming region and with low sigmas
in $U$, $V$ and $W$ are dispersed during their rotation around the Galactic centre,
inducing a particular shape in both the space and the velocity field since some of
the stars fall behind while others go ahead. The Galactic potential maintains
the group bounded during several hundreds of years, in spite of the initial velocity
dispersion in the molecular cloud, in both the $UV$-plane and the $W$ component.

\section{The AB Dor MG and subgroup B4}
\label{sec:abdor}

Very recently, \citet{zuc04} have identified a large group of stars with the
same space motion than the well-known young K-dwarf AB~Dor (d = 15~pc),
a quadruple system \citep{clo05, gui05} made up of three late-type stars
--- AB~Dor~A (HD~36705),
AB~Dor~Ba and AB~Dor~Bb --- and a very low mass companion which has recently
been object of discussion because of the discrepancy between its
dynamical mass and that predicted by evolutionary models
\citep{clo05}.
All the stars listed in Table~1 of \citet{zuc04} are situated inside
the Local Association (see Fig.~\ref{fig:uv}) near the boundaries of the young
disk stellar population \citep{egg84},
and have at least one indicator of
youth. Taking the intensity of the H$\alpha$ emission line of
these stars and the position in a $V-K_{\rm s}$
diagram of three M-type members of the MG into account,
they estimate an age of $50\pm10$ Myr for the AB~Dor MG.
%
Very recently, \citet{luh05} and \citet{luh05b}
have showed that the components of
AB~Dor should have an age of 75 -- 150 Myr based on the comparison of both
their position in the $M_{\rm K}$ vs. $V-K_{\rm s}$ diagram
with respect to the Pleiades and IC~2391 clusters,
and the $EW$(Li~{\sc i}) of AB~Dor~A with that of rapidly
rotating K dwarfs in the Pleiades. Moreover, with an age of
$\sim$~100 Myr the discrepancy between observations and models
for the very low-mass companion (AB~Dor~C)
would disappear \citep[eg.][]{clo05}. Taking this into account,
they propose an age range of 75 -- 150 Myr for all the MG.

To the initial sample of \citet{zuc04}, we have added 13 stars (marked with
{\it b} in Table~\ref{tab:MGs}) from our catalogue of {\it Late-type Stars
Possible Members of Young Stellar Kinematic Groups} \citep{m01}.
These stars have been included to both searches: a) for other members
of the group and b) to show the existence of two subgroups of different
ages more clearly.
They have been chosen because of their kinematics, assuming
a total dispersion of \mbox{$\pm 4$ km s$^{-1}$} in $U$ and $V$ respectively,
around the centre of the AB~Dor~MG defined by \citet{zuc04}.
We have imposed no restriction to the $W$ component for this first
selection. The whole sample contains a total of 50 stars.

We have compared the $EW$({Li~{\sc i}}) of every star in the sample with
that of known members of young open clusters (Fig.~\ref{fig:li_MGs}), as well
as their position in the $V-I$ diagram with the isochrones of
\citet{sie00} (Fig.~\ref{fig:VI}). The results reveal the existence
of two subgroups with stars showing different spectroscopic and photometric
features,
mixed in the velocity field (see Fig.~\ref{fig:uvw_MGs}).
The members of the first subgroup, that includes AB~Dor and
PW~And --- a very active young K2-dwarf \citep{lop03} --- show
$EW$({Li~{\sc i}}) similar to that of the high-rotators in the Pleiades
(upper continuous line in Fig.~\ref{fig:li_MGs}) which are
above the values found in the low-rotators of
IC~2602 (lower dashed line in Fig.~\ref{fig:li_MGs}).
Their position between the 30 and the 80 Myr isochrones
in the $V-I$ diagram, together with the first result,
is compatible with an age of 30 - 50~Myr. Moreover, the stars from the
sample of \citet{zuc04} belonging to this subgroup are
situated above the sequence of the Pleiades in the $M_{\rm K}$ vs.
$V-K_{\rm s}$ diagram in \citet{luh05}.
Here we have obtained a dispersion $\sigma \approx 2$ km~s$^{-1}$ in the
W component, quite similar to the one observed in other young stellar
associations such as Tucana or $\epsilon$~Cha
\citep[see][and references therein]{zuc04a}.
For determining the dispersion we have rejected the stars BD+07~1919A and
B (marked with * in Table~\ref{tab:MGs}) since their radial velocities ---
used for calculating the Galactic velocity components --- have not been
corrected for binarity since no orbital solution has been found in the
literature. Nevertheless, although the membership of this system is not
completely reliable taking the value of their $W$ component into account,
it has been included in the sample as possible member because of the
position of the A component in the CMD, which suggests an age of
$\sim$~30~Myr.
The stars in the second subgroup show features, in terms of
$EW$({Li~{\sc i}}) values and position in the CMD,
comparable with that of the members of the Pleiades cluster:
in Fig.~\ref{fig:li_MGs} they are situated slightly above the
lower envelope of the Pleiades, while in Fig.~\ref{fig:VI}
they situate on the (ZAMS) 80~Myrs isochrone.
Its members could be considered as part of
subgroup B4, one of the four subgroups found by \citet{asi99}
inside the Local Association in their study of the space motion
of OB~associations using Hypparcos astrometric data. The
authors find a mean age of $\sim$~150~Myr for this subgroup
using information from the photometry.
The higher dispersion found for the stars of this second subgroup in the
velocity space (Fig.~\ref{fig:uvw_MGs}) is in agreement with the age
estimated by us.

On the other hand,
the results about AB~Dor MG indicate that this quadruple system
has indeed $\sim$~50~Myr.
The value of $EW$(Li~{{\sc i}}) for AB~Dor~A is somewhat above the upper
envelope of the Pleiades but not so high as the one of IC~2602
(Fig.~\ref{fig:li_MGs}). On the other hand, its ($V-I$) colour situates
it between the 30 and 80 Myr isochrones (Fig.~\ref{fig:VI}). The same
result is clearly visible in Fig.~1 of \citet{luh05}, where AB~Dor
is situated above the lower sequence of the Pleiades in the
$M_{\rm K}$ vs. $V-K_{\rm s}$ diagram. With an age of 50~Myr,
the discrepancy between observations and models
for the AB Dor very low-mass companion (AB Dor C)
shown in \citet{clo05} continuous, although it can be solved
if the very low-mass companion were indeed an
unresolved binary system \citep{mar05}.


\section{Discussion and conclusions}
\label{sec:conc}

In Table~\ref{tab:MGs} we list the stars belonging to the nearest moving
groups: Hercules-Lyra Association and AB~Dor MG, and those being part of
the Local Association B4 subgroup.
For the Hercules-Lyra Association, a division between certain members and
candidates with doubtful classification or non members has been made.
In the three groups, new candidates from our catalogue of {\it Late-type Stars
Possible Members of Young Stellar Kinematic Groups} \citep{m01} have been
selected because of their kinematics (see $\S$~\ref{sec:herc}
and $\S$~\ref{sec:abdor}). A total of
75 stars including the known members and the new candidates selected by us
have been analysed. Kinematic, spectroscopic and photometric criteria have
been utilized to discriminate non members from the rest of candidates of
the Hercules-Lyra Association, and to distinguish between the members of
the AB~Dor MG and those of the B4 subgroup.

\begin{figure}[!t]
\epsscale{1.0} \plotone{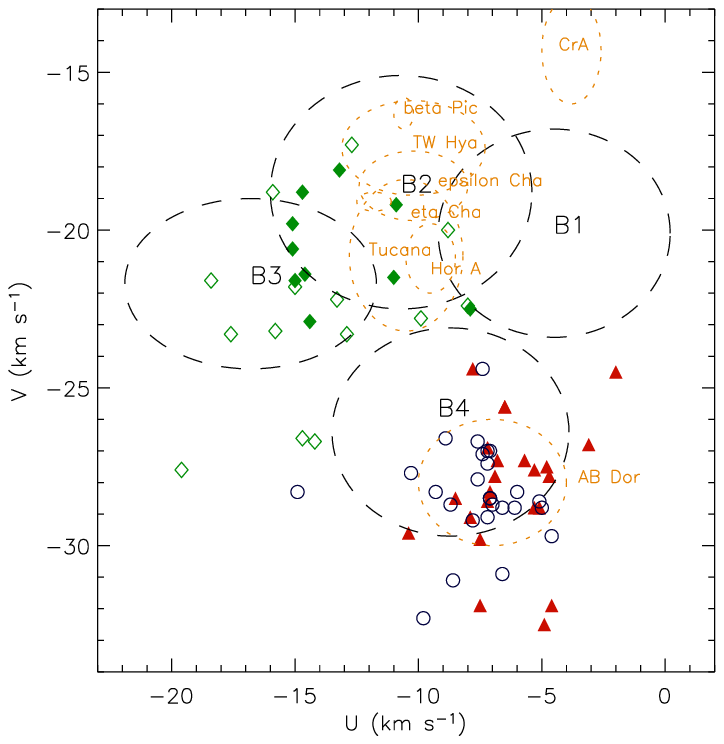}
\caption{Position in the UV-plane of the
stars listed in Table~\ref{tab:MGs} and young stellar associations.
Symbols are used as follows: filled diamonds for Hercules-Lyra members;
open diamonds for non members of Hercules-Lyra or stars with doubtful
classification; triangles for AB~Dor MG; and circles for other Local
Association members (members of the subgroup B4).
%
\label{fig:uv}}
\end{figure}

\begin{figure}[!t]
\epsscale{1.0} \plotone{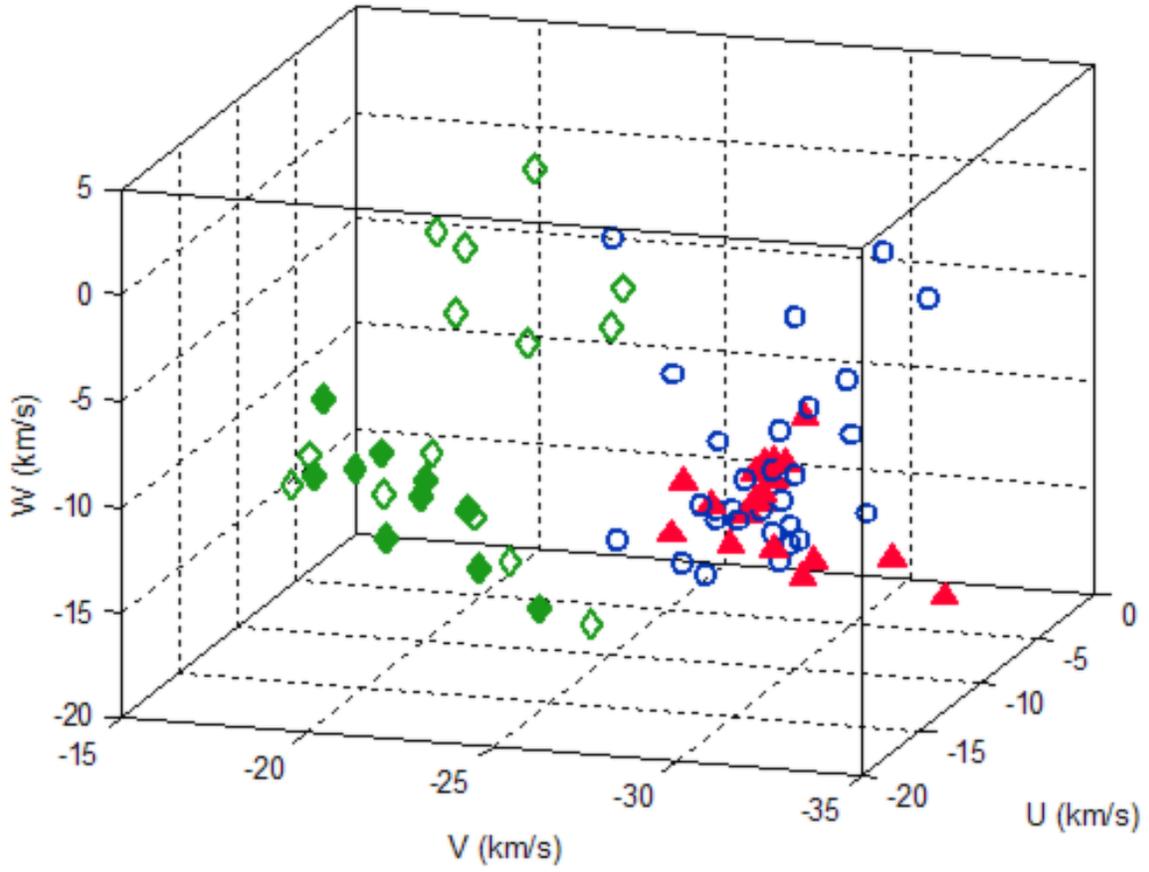}
\caption{Position in the 3D velocity space
of the stars in Fig.~\ref{fig:uv}. Candidates of Hercules-Lyra with values
of $W$ different from that of the rest of the group are clearly
distinguishable. \label{fig:uvw_MGs}}
\end{figure}

\begin{figure}[!t]
\plotone{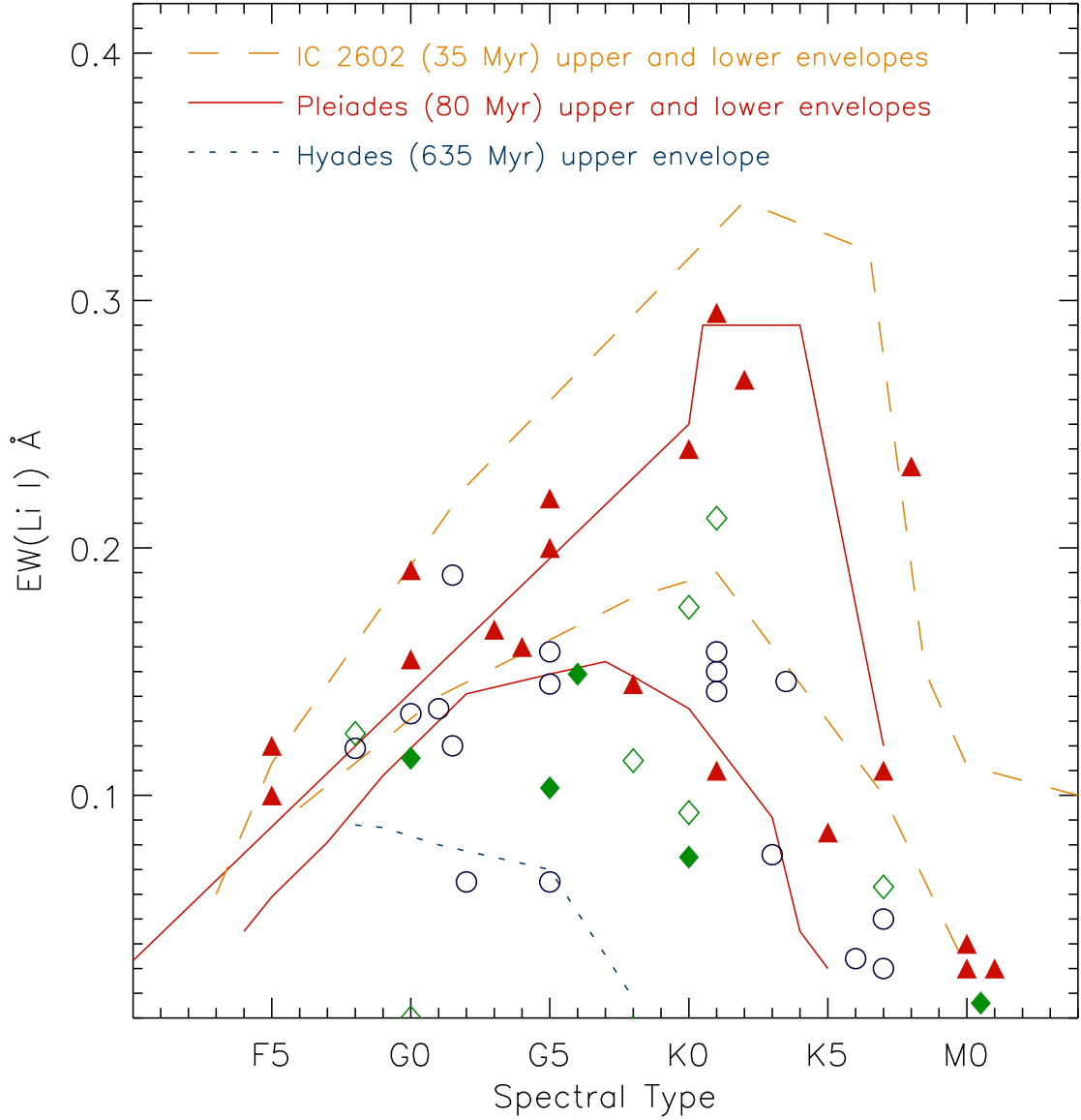}
\caption{Equivalent width of Li~{\sc i} 6707.8~\AA \ as
a function of spectral type for the stars in Table~\ref{tab:MGs}, compared
with the envelopes of well-known stellar clusters. Symbols as in
Fig.~\ref{fig:uv}.
%
\label{fig:li_MGs}}
\end{figure}

\begin{figure}[!t]
\plotone{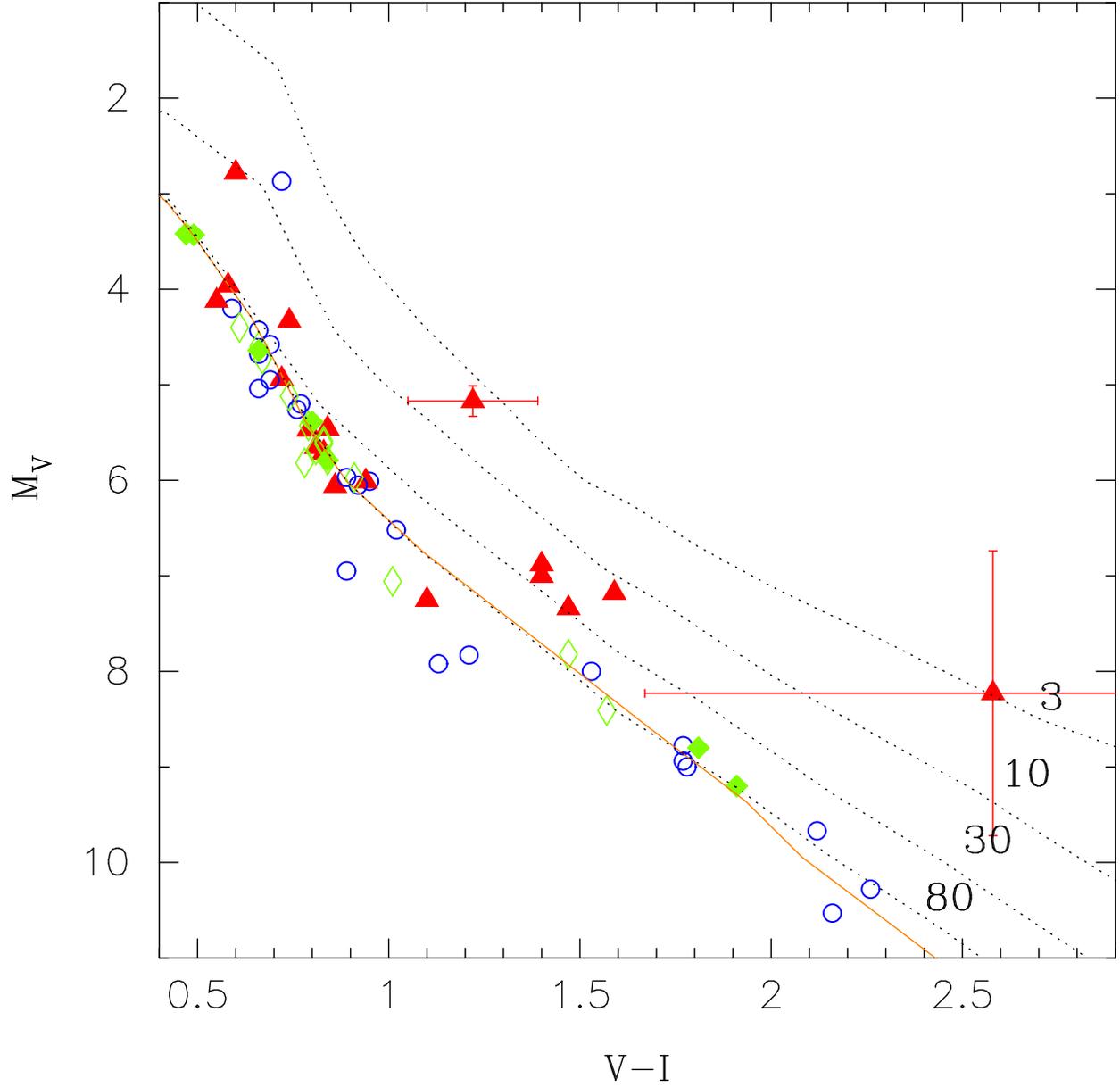}
\caption{$M_{\rm V}$ vs. $V - I$ for the stars of the
Hercules-Lyra Association, AB~Dor MG and Local Association subgroup B4.
Symbols as in Fig.~\ref{fig:uv}. The member of AB~Dor MG situated under
the ZAMS is LO~Peg, which photometry is probably affected by dark
spot-type features. Isochrones of 3, 10, 30 and 80 Myr from \citet{sie00}
are plotted as well as ZAMS (continuous line).
\label{fig:VI}}
\end{figure}

In the velocity space, Hercules-Lyra is clearly distinguishable from
the rest of the sample (see Figs.~\ref{fig:uv} and~\ref{fig:uvw_MGs}).
The dispersion in $U$, $V$, and $W$ is comparable with that of other
coeval MGs such as Castor and Ursa Major \citep[e.g.][]{m01}, and
compatible with its age (see $\S$~\ref{sec:herc}).
On the other hand, AB~Dor~MG and B4 subgroup are mixed up and age-dating
criteria are necessary to distinguish between the members of both groups.
Nevertheless, the dispersion in $W$ for AB~Dor MG is quite smaller than
the one of B4 subgroup.
Age-dating criteria are also necessary to discriminate non members of
Hercules-Lyra from the certain ones. The results of applying them are
summarized in Table~\ref{tab:MGs}: the Hercules-Lyra Association is formed
by 10 certain members
situated at a mean distance of $\sim$~20~pc and show values of
$EW$(Li~{{\sc i}}) (Fig.~\ref{fig:li_MGs}) and a position in the $V-I$
CMD (Fig.~\ref{fig:VI}) compatible with an age of 150~--~300~Myr;
the members of AB~Dor MG are situated at a mean distance of $\sim$~30~pc and
show lithium abundances typical of stars with 30~--~50~Myr
(Fig.~\ref{fig:li_MGs}), which is in
agreement with their position in the $M_{\rm V}$ vs. $V-I$ diagram
(Fig.~\ref{fig:VI});
finally, a set of stars with $EW$(Li~{{\sc i}}) and positions in the
CMD compatible with an age of 80~--~120~Myr are
mixed with Hercules-Lyra and AB~Dor MG, and have been classified as
other members of the Local Association B4 subgroup (see $\S$~3).
Note that the age estimated using the position of the members of
Hercules-Lyra in the CMD
is a lower limit since the 80~Myrs isochrone is overlapped with the
ZAMS for spectral types earlier than about K5. On the other hand,
the age estimated using the equivalent width of the Li~{\sc i} line
$\lambda$6707.8 \AA \ is more robust
since the 50\% of the stars classified as members have measurements of
the $EW$(Li~{{\sc i}}): the Li indicator is useful only for
spectral type later than G0, but only three of the 25 candidates of the
initial sample are F stars.

Stars in these three subgroups form an excellent list of young cool
stars for studying how planets are formed, since they cover a range of ages
between 30 and 200 Myr, characteristic of the period during which the
Solar System was formed, and they are close enough to be accessible to
adaptive optics.
In addition, they can be taken as targets for direct imaging detection
of sub-stellar companions ---~brown dwarfs and extra-solar giant planets~---
\citep{neu00,mar03,mas05,low05} and for cold dust, debris disks
\citep{gai04,met04,liu04,che05}.

\acknowledgments

This work was supported by the Universidad Complutense de Madrid and
the Spanish Ministerio de Educaci\'on y Ciencia (MEC), Programa
Nacional de Astronom\'{\i}a y Astrof\'{\i}sica under grants AYA2004-03749
and AYA2005-02750.
ICC acknowledges support from MEC under AP2001-0475.
We would like to thanks the referee for useful comments
which have contributed to improve the manuscript.

\clearpage




\clearpage

\clearpage





\clearpage

\clearpage

\clearpage

\clearpage

\begin{deluxetable}{lccccccccccc}
\setlength{\tabcolsep}{0.02in}
\tabletypesize{\tiny}
\tablecaption{Stars members of near MGs.\label{tab:MGs}}
\tablehead{
\colhead{HD/} &
\colhead{R.A.} &
\colhead{Dec} &
\colhead{SpT} &
\colhead{$D$} &
\colhead{$V_{\rm hel} \pm \sigma_{\rm V_{\rm hel}}$} &
\colhead{$U$} &
\colhead{$V$} &
\colhead{$W$} &
\colhead{$B-V$} &
\colhead{$V-I$} &
\colhead{$EW$(Li~{{\sc i}})$^{c}$} \\
\colhead{other name} &
\colhead{(J2000.0)} &
\colhead{(J2000.0)} &
\colhead{} &
\colhead{(pc)} &
\colhead{(km s$^{-1}$)} &
\colhead{(km s$^{-1}$)} &
\colhead{(km s$^{-1}$)} &
\colhead{(km s$^{-1}$)} &
\colhead{(mag)} &
\colhead{(mag)} &
\colhead{(m\AA)}
}
\startdata
\hline
\multicolumn{4}{l}{\bf Hercules-Lyra Association: members} \\
\hline
166$^\dag$  & 00 06 36.78  &~29 01 17.41  & K0    V      & 13.7 &~-6.9  0.2& -15.0 & -21.6 &-10.0 & 0.75 & 0.80 & ~75\\
10008   & 01 37 35.47  &-06 45 37.52  & G5    V      & 23.6 &~11.6  0.6& -13.2 & -18.1 &-11.1 & 0.80 & 0.84 & 103\\
233153$^a$  & 05 41 30.73  &~53 29 23.28  & M0.5         & 12.5 &~~1.9  1.0& -14.4 & -22.9 &-14.3 & 1.40 & 1.91 & ~16\\
HIP 37288$^a$& 07 39 23.04  &~02 11 01.18  & K7           & 14.9 &~18.5  5.0& -11.0 & -21.5 &-13.1 & 1.38 & 1.81 & ~~-  \\
70573$^a$   & 08 22 49.95  &~01 51 33.55  & G6    V      & 45.7 &~19.5  1.0& -14.7 & -18.8 &~-6.7 & 0.59 &   -  & 149\\
HIP 53020$^a$& 10 50 52.06  &~06 48 29.34  & M4           & 12.9 &~-2.0  0.1& ~-7.9 & -22.5 &-19.1 & 1.68 & 2.81 & ~~- \\
GJ 560B$^a$ & 14 42 30.42  &-64 58 30.50  & K5    V      & 16.4 & ~~7.0  4.0 & -10.9 &-19.2 &-10.8 & 1.15 &   -  & ~~- \\
139664$^a$  & 15 41 11.38  &-44 39 40.34  & F5    V      & 17.5 &~-5.4  2.0& -15.1 & -19.8 &~-9.7 & 0.41 & 0.47 & ~~- \\
206860$^\dag$& 21 44 31.33  &~14 46 18.98  & G0    V      & 18.4 &-16.9  2.0& -14.6 & -21.4 &-11.0 & 0.58 & 0.66 & 115\\
213845$^a$  & 22 34 41.64  &-20 42 29.56  & F7    V      & 22.7 &~-1.9  0.9& -15.1 & -20.6 &-12.9 & 0.45 & 0.49 & ~~- \\
\hline
\multicolumn{7}{l}{\bf Hercules-Lyra Association: non members or doubtful classification} \\
\hline
1466$^a$    & 00 18 26.12  &-63 28 38.97  & F8    V      & 40.9 & 0.54  2.0& ~-8.8 & -20.0 &~-1.2 & 0.54 & 0.61 & 125\\
17925       & 02 52 32.13  &-12 46 10.97  & K1    V      & 10.4 &~17.5  0.1& -15.0 & -21.8 &~-8.7 & 0.88 & 0.91 & 212\\
1E 0318.5-19.4$^a$&03 20 49.50  &-19 16 10.00  & K7    V      & 27.0 &~20.8  1.0& -12.7 & -17.3 &-11.8 &    - &    - & ~63\\
37394       & 05 41 20.34  &~53 28 51.81  & K1    V      & 12.2 &~~0.3  0.2& -12.9 & -23.3 &-14.5 & 0.84 & 0.88 & ~~2\\
82443$^\dag$& 09 32 43.76  &~26 59 18.71  & K0    V      & 17.7 &~~8.1  0.1& ~-9.9 & -22.8 &~-5.6 & 0.78 & 0.78 & 176\\
96064       & 11 04 41.47  &-04 13 15.91  & G8    V      & 24.4 &~18.3  0.8& -14.2 & -26.7 &~-0.6 & 0.77 & 0.81 & 114\\
97334B$^\dag$& 11 12 32.35  &~35 48 50.69  & G0    V      & 21.7 &~-3.6  1.0& -15.8 & -23.2 &-11.2 & 0.60 & 0.67 & ~10\\
111395      & 12 48 47.05  &~24 50 24.81  & G5    V      & 17.7 &~-8.6  1.0& -18.4 & -21.6 &~-9.2 & 0.70 & 0.74 & ~~0\\
112733$^a$  & 12 58 31.97  &~38 16 43.55  & K0    V      & 22.5 &~-3.4  0.1& -17.6 & -23.3 &~-0.8 & 0.74 & 0.79 & ~93\\
116956      & 13 25 45.53  &~56 58 13.77  & G9    V      & 21.8 &-13.1  0.3& -15.9 & -18.8 &~-8.8 & 0.80 & 0.83 & ~~0\\
HIP 67092$^a$& 13 45 05.33  &-04 37 13.25  & K5           & 25.7 &~~4.6  0.5& ~-8.0 & -22.4 &~~1.8 & 1.49 & 1.57 & ~~- \\
139777      & 15 29 11.20  &~80 26 55.00  & F0    V      & 22.1 &-15.8  0.5& -14.7 & -26.6 &~-2.2 &    - &    - & ~~- \\
139813$^\ddag$& 15 29 23.60  &~80 27 01.00  & G5    V      & 21.7 &-15.8  0.5& -14.7 & -26.6 &~-2.2 & 0.80 & 0.83 & ~~- \\
141272      & 15 48 09.46  &~01 34 18.26  & G8    V      & 21.4 &-27.2  0.3& -19.6 & -27.6 &-14.0 & 0.80 & 0.84 & ~~6\\
207129$^a$  & 21 48 15.75  &-47 18 13.01  & G0    V      & 15.6 &~-6.5  1.3& -13.3 & -22.2 &~~0.3 & 0.60 & 0.66 & ~~- \\
\hline
\multicolumn{2}{l}{\bf AB Doradus Moving Group} \\
\hline
1405        & 00 18 20.90  &~30 57 22.03  & K2    V      & 30.6 & -11.2  0.1 & ~-5.3 &-28.8 &-17.8 & 1.04 &    - & 268 \\
HIP 6276    & 01 20 32.27  &-11 28 03.74  & (G8)         & 35.1 & ~~9.9  1.0 & ~-4.7 &-27.8 &-13.6 & 0.79 & 0.83 & 145 \\
13482       & 02 12 15.41  &~23 57 29.54  & K1 + K5      & 32.3 & ~-1.3  0.3 & ~-7.1 &-28.3 &-11.8 & 1.13 & 1.22 & 110 \\
17332       & 02 47 27.42  &~19 22 18.56  & G0 + G5      & 32.6 & ~~4.1  1.3 & ~-8.5 &-28.5 &-12.9 & 0.68 & 0.74 & 155 \\
19668$^b$   & 03 09 42.29  &-09 34 46.59  & G0    V      & 40.2 & ~14.6  0.7 & ~-5.1 &-28.8 &-10.3 & 0.81 & 0.84 & 191 \\
21845       & 03 33 13.49  &~46 15 26.54  & (G5)         & 33.8 & ~-6.0  0.3 & ~-6.5 &-25.6 &-15.7 & 0.70 & 0.81 & 200 \\
HIP 16563B  & 03 33 14.00  &~46 15 19.00  & M0           & 33.8 & ~-6.1  1.1 & ~-6.5 &-25.6 &-15.7 &   -  &   -  & ~30 \\
25457       & 04 02 36.74  &-00 16 08.12  & F5 V         & 19.2 & ~17.0  0.3 & ~-7.2 &-28.6 &-11.6 & 0.52 & 0.58 & 100 \\
25953       & 04 06 41.53  &~01 41 02.08  & F5           & 55.3 & ~17.6  0.6 & ~-6.9 &-27.8 &-14.3 & 0.48 & 0.55 & 120 \\
36705       & 05 28 44.83  &-65 26 54.85  & K1           & 14.9 & ~33.0  3.0 & ~-7.5 &-29.8 &-16.0 & 0.83 & 0.94 & 295 \\
37572       & 05 36 56.85  &-47 57 52.87  & K0 V         & 23.9 & ~31.0  1.0 & ~-7.2 &-26.9 &-13.9 & 0.84 & 0.86 & 240 \\
BD+20 1790$^b$& 07 23 44.00  &~20 25 06.00  & K5    V      & 31.6 & ~19.9  0.1 & ~-4.9 &-32.5 &-18.5 & 1.07 &    - & ~85 \\
89744$^b$   & 10 22 10.56  &~41 13 46.31  & F7    V      & 39.0 & ~-6.5  1.3 & -10.4 &-29.6 &-14.2 & 0.53 & 0.60 & ~~- \\
139751      & 15 40 28.39  &-18 41 46.19  & (K7)         & 42.6 & ~-8.9  0.4 & ~-7.5 &-31.9 &-15.6 & 1.24 & 1.40 & 110 \\
160934      & 17 38 39.63  &~61 14 16.12  & M0           & 24.0 & -35.6  0.7 & ~-5.3 &-27.6 &-14.5 & 1.30 & 2.58 & ~40 \\
HIP 106231  & 21 31 01.71  &~23 20 07.37  & K8           & 25.1 &            & ~-5.7 &-27.3 &-15.0 & 1.03 & 1.10 & 233 \\
217343      & 23 00 19.29  &-26 09 13.50  & G3 V         & 32.1 & ~~6.3  1.5 & ~-3.1 &-26.8 &-14.1 & 0.65 & 0.72 & 167 \\
217379      & 23 00 27.96  &-26 18 42.80  & (K8)         & 30.0 & ~~8.4  1.5 & ~-2.0 &-24.5 &-15.4 & 1.34 & 1.59 & ~~0 \\
HIP 114066  & 23 06 04.84  &~63 55 34.36  & (M1)         & 24.9 & -23.7  0.8 & ~-6.8 &-27.3 &-15.9 & 1.21 & 1.77 & ~30 \\
218860      & 23 11 52.05  &-45 08 10.63  & (G5)         & 50.5 & ~10.3  1.2 & ~-7.9 &-29.1 &-11.3 & 0.71 & 0.76 & 220 \\
HIP 115162  & 23 19 39.56  &~42 15 09.82  & (G4)         & 49.4 & -19.7  0.2 & ~-4.8 &-27.5 &-14.3 & 0.75 & 0.79 & 160 \\
\hline
\multicolumn{2}{l}{\bf Members of subgroup B4} \\
\hline
4277        & 00 45 50.89  &~54 58 40.17  & F8 V + K3    & 48.5 & -15.4  0.5 & ~-8.9 &-26.6 &-15.8 & 0.52 & 0.59 & 119 \\
6569        & 01 06 26.15  &-14 17 47.11  & K1 V         & 50.0 & ~~6.0  1.2 & ~-8.6 &-31.1 &~-9.3 & 0.91 & 0.95 & 150 \\
HIP 12635   & 02 42 20.97  &~38 37 21.20  & (K3.5)       & 49.6 & ~-4.1  0.3 & ~-8.7 &-28.7 &-13.1 & 0.88 & 0.89 & 146 \\
16760       & 02 42 21.31  &~38 37 07.20  & G5           & 49.6 & ~-3.3  0.2 & ~-9.3 &-28.3 &-13.4 & 0.71 & 0.77 & 158 \\
HIP 14807   & 03 11 12.34  &~22 25 22.77  & (K6)         & 49.8 & ~~4.1  0.3 & ~-5.1 &-28.6 &-16.1 &   -  &   -  & ~34 \\
HIP 14809   & 03 11 13.84  &~22 24 57.11  & G5           & 49.8 & ~~5.2  0.2 & ~-6.0 &-28.3 &-16.7 & 0.71 & 0.66 & 145 \\
HIP 17695   & 03 47 23.35  &-01 58 19.93  & M3           & 16.3 & ~16.0  1.7 & ~-7.4 &-27.1 &-10.6 & 1.51 & 2.16 & ~~-\\
25457       & 04 02 36.74  &~00 16 08.13  & F6    V      & 19.2 & ~15.3  0.4 & ~-6.0 &-28.3 &-10.5 & 0.52 & 0.58 & 118\\
35650       & 05 24 30.17  &-38 58 10.76  & (K7)         & 17.7 & ~30.9  1.0 & ~-7.1 &-27.0 &-14.5 & 1.25 & 1.21 & ~~0 \\
HIP 26369   & 05 36 55.07  &-47 57 47.99  & (K7)         & 23.9 & ~31.1  1.1 & ~-7.2 &-27.0 &-13.9 & 1.17 & 1.13 & ~30 \\
45270       & 06 22 30.94  &-60 13 07.15  & G1 V         & 23.5 & ~30.0  0.7 & ~-7.6 &-26.7 &-13.6 & 0.61 & 0.66 & 135 \\
GSC 8894-426& 06 25 55.39  &-60 03 29.20  & M2           & (22) & ~31.8  2.0 & -10.3 &-27.7 &-15.6 &   -  &   -  & ~~0 \\
48189       & 06 38 00.36  &-61 32 00.19  & G1.5 V       & 21.7 & ~33.4  1.0 & ~-7.1 &-28.5 &-15.0 & 0.62 & 0.69 & 120 \\
HIP 31878   & 06 39 50.02  &-61 28 41.52  & (K7)         & 21.9 & ~30.5  0.7 & ~-7.2 &-27.4 &-13.9 & 1.26 & 1.53 & ~50 \\
HIP 36349$^b$& 07 28 51.37  &-30 14 48.54  & (M3)         & 15.6 & ~26.6  1.0 & ~-7.4 &-24.4 &-15.7 & 1.44 & 1.78 & ~~0 \\
BD+07 1919A$^{b,*}$& 08 07 09.09  &~07 23 00.13  & K5    V      & 40.2 & ~19.1 0.1 & ~-4.6 &-31.9 &~-4.5 & 1.24 & 1.40 & ~~- \\
BD+07 1919B$^{b,*}$& 08 07 08.78  &~07 22 58.39  & K7    V      & 40.2 & ~18.7 0.1 & ~-7.8 &-24.4 &~-1.2 & 1.15 & - & ~~- \\
HIP 51317$^b$& 10 28 55.55  &~00 50 27.58  & M2           & 26.1 & ~~8.3  0.5 & ~-7.8 &-29.2 &-15.1 & 1.50 & 2.26 & ~~- \\
92945$^b$   & 10 43 28.27  &-29 03 51.43  & K1    V      & 21.6 & ~23.2  0.6 & -14.9 &-28.3 &~-4.0 & 0.87 & 0.92 & 158 \\
GJ 466$^b$  & 12 25 58.58  &~08 03 44.03  & M0    V      & 39.9 & ~10.0  0.1 & ~-9.8 &-32.3 &~~0.1 & 1.46 & 1.47 & ~~- \\
113449      & 13 03 49.66  &-05 09 42.52  & (K1)         & 22.1 & ~~2.0  0.5 & ~-5.0 &-28.8 &~-9.8 & 0.85 & 0.89 & 142 \\
129333$^b$  & 14 39 00.22  &~64 17 29.84  & G1.5  V      & 33.9 & -20.6  0.3 & ~-7.2 &-29.1 &~-4.6 & 0.63 & 0.69 & 189 \\
HIP 81084   & 16 33 41.61  &-09 33 11.95  & M0.5         & 31.9 & -15.0  0.4 & ~-7.0 &-28.7 &-13.4 & 1.44 & 1.77 & ~~0 \\
152555      & 16 54 08.14  &-04 20 24.66  & G0           & 47.6 & -17.1  0.5 & ~-6.1 &-28.8 &-12.6 & 0.59 & 0.66 & 133 \\
199065A$^b$ & 20 57 22.44  &-59 04 33.46  & G2    V      & 50.9 & ~11.0  2.0 & ~-7.1 &-28.5 &-12.0 & 0.66 & 0.72 & ~65 \\
199065B$^b$ & 20 57 21.86  &-59 04 34.23  & G5    V      & 50.9 & ~11.0  2.0 & ~-4.6 &-29.7 &~-8.6 &   -  &   -  & ~65 \\
GJ 856A     & 22 23 29.09  &~32 27 33.47  & M0    V      & 16.1 & -24.0  3.0 & ~-6.6 &-30.9 &-13.9 & 1.57 & 2.12 & ~~- \\
GJ 856B$^b$ & 22 23 30.00  &~32 27 00.00  & M1    V      & 16.1 & -21.7  1.0 & ~-6.6 &-28.8 &-14.7 & 1.49 &    - & ~~- \\
224228      & 23 56 10.67  &-39 03 08.40  & K3 V         & 22.1 & ~12.1  0.5 & ~-7.6 &-27.9 &-12.3 & 0.97 & 1.02 & ~76 \\
\hline
\enddata
\tablecomments{$^\dag$ Co-moving group \citep{gai98}.
$^\ddag$ $V_{\rm r}$ and ($U$, $V$, $W$) from the A component (HD~139777).
$^a$ New candidates of Hercules-Lyra Association.
$^b$ Stars added to the initial sample of \citet{zuc04}.
$^c$ $EW$(Li~{\sc i}) from \citet{lop05} and \citet{lop05b} for the stars in
the Hercules-Lyra Association, for the stars marked with $^b$ in AB~Dor
MG and B4 subgroup and for HD~1405 (PW~And) and HIP~106231 (LO~Peg).
For the rest of stars we have adopted the values of \citet{zuc04}.
$^*$ Doubtful members of the AB~Dor MG.}
\end{deluxetable}

\end{document}